\documentstyle[prl,twocolumn,aps]{revtex}

\begin{document}
\draft

\title{Bound entanglement can be activated}

\author{Pawe\l{} Horodecki\cite{poczta1}}

\address{Faculty of Applied Physics and Mathematics\\
Technical University of Gda\'nsk, 80--952 Gda\'nsk, Poland}

\author{Micha\l{} Horodecki\cite{poczta2} and Ryszard Horodecki\cite{poczta3}}

\address{Institute of Theoretical Physics and Astrophysics\\
University of Gda\'nsk, 80--952 Gda\'nsk, Poland}

\maketitle

\begin{abstract}
Bound entanglement is the noisy entanglement
which cannot be distilled to a singlet form.
Thus it cannot be used alone for quantum communication
purposes. Here we show that, nevertheless,
the bound entanglement can be, in a sense,
pumped into single pair of free entangled particles.
It allows for teleportation via the pair with the fidelity
impossible to achieve without support of bound entanglement.
The result also suggests that the distillable
entanglement may be not additive.
\end{abstract}
\pacs{Pacs Numbers: 03.65.Bz}

Despite deep research, quantum entanglement
still astonishes even specialists, producing highly
nonintuitive effects
such as quantum paralelism \cite{Deusch},
quantum cryprography \cite{Ekert}, quantum dense coding \cite{geste},
quantum teleportation \cite{Bennett_tel},
reduction of communication complexity \cite{compl}.
In practice, one usually deals with noisy
entanglement represented
by {\it mixed} states of composite system.
The latter are entangled (inseparable) if they are not mixtures of
product states
\cite{Werner,tran}.
However, the mixed state entanglement cannot be
used directly for quantum communication purposes.
For this reason the first example procedure of distillation
of it to useful singlet form represented by the two spin-${1 \over 2}$
singlet state
$\Psi_-={1\over \sqrt2} (|\uparrow\downarrow\rangle-|\downarrow\uparrow\rangle)$
has been provided by Bennet {\it et. al} \cite{d1}
and discussed later in \cite{huge}.
Similar procedure has been applied to quantum privacy
amplification \cite{Deutsch}.
Subsequently, it has been shown \cite{my1} that, noisy entanglement of two
spin-${1 \over 2}$ system, however small, can be distilled to a the singlet
form. Then it was naturally supposed that the same  is possible for larger
systems. However, quite recently, it has been shown that
begining with two spin-1 systems, quantum mechanics
implies existence of two qualitatively different
kinds of noisy entanglement \cite{bound}:
apart from
the ``free''  entanglement which is distillable there is a ``bound''
one which by no means can be brought to the singlet form.
The curiosity of the bound entangled states is that to produce them one
needs some amount of pure entanglement, while any, however little amount of it
cannot be recovered back from them.
The bound entanglement is  closely connected with Peres separability test
\cite{Peres} (see also \cite{sep}).
In particular, it has been shown \cite{bound}
that if an inseparable state satisfies Peres criterion, then its
entanglement is bound.

The existence of bound entanglement involves new
questions concerning local realism and quantum information.
However there is a question closely related to the practical topics.
Namely one can simply ask: Can the bound entanglement be somehow
activated to produce {\it any} effect useful in quantum communication?
In this paper we show that the bound entanglement can be, in a sense,
liberated, giving, in particular, some chance of improving the transmission
of quantum information and suggesting existence of qualititatively new
processes in mixed entanglement domain.

Before we state the main results of this paper let us recall
that quite recently \cite{Linden,Kent} it has been pointed out
that mixed free entanglement may have some disavantage
as it cannot be distilled {\it noncollectively}
i.e. by acting over any given pair of particles separately.
It particular it means that in some cases
given {\it single} pair of two spin-s particles in free entangled (FE)
state $\varrho_{in}$, using only quantum local operations (QL) and
classical communication (CC), cannot make the {\it fidelity}
of the resulting state $\varrho_{out}$
\begin{eqnarray}
F(\varrho)=\langle\Psi_+|\varrho|\Psi_+\rangle , \ \ \
|\Psi_{+} \rangle ={1\over \sqrt{2s+1}}\sum_{i=0}^{2s} |i \rangle | i \rangle
\label{fid}
\end{eqnarray}
arbitrary close to 1.
This is an important point as quantity (\ref{fid}),
measuring how close is $\varrho$ to maximally entangled two spin-s state
$\Psi_{+}$,
plays central role in the teleportation scheme \cite{Bennett_tel}
if applied to mixed states \cite{Pop}.

Now, let us explain the main result of
this paper. We consider just a {\it single}
pair of spin-1 particles in a mixed state $\varrho$
shared by paradigmatic, spatially separated
Alice and Bob who are allowed to make any QLCC operations.
The state $\varrho$ is taken to be free entangled (FE), but
its entanglement is chosen to be so weak that
in the case of single pair no QLCC operations can increase its fidelity
upon some bound $C<1$. We then introduce some new bound entangled (BE)
states and show that if, in addition,
Alice and Bob are provided with a large supply of pairs
in those states, then they can skip the border $C$
making now the fidelity of original FE pair
arbitrary close to 1 with nonzero probability.
We shall hereafter call the process of making the fidelity
$F$ arbitrary close to unity
{\it quasi-distillation}, as, in contrast with the original distillation idea,
we allow the number of initial pairs and the probability of
success to depend on the required final $F$. 
The key point of the presented result is
here that the distinguished FE pair as well
as the set of all BE pairs cannot be quasi-distilled themselves.
However, putting them together produces new
quality from which the state with arbitrary good
fidelity already can be obtained. The revealed
process can be viewed as a kind of entanglement transfer from BE
pairs into FE pair.
After the presentation of details of the effect (see below) we adress
the question of possible relevance of the effect for quantum
communication and show that some transfer which
is impossible with FE pair alone sometimes
can be done with aid of bound entanglement supply.
This supports the hope for some usefulness
of bound entanglement for quantum communication purposes.
Finally  we discuss possible relevance of the effect
in the context of the original distillation idea. In particular we conclude
that, in the present context, it can not be excluded that the
distillable entanglement \cite{huge} is not additive.

To ilustrate details of scheme consider the case of two spin-1 particles.
The state of any particle can be described using the
three-dimensional Hilbert space spanned by the basis
states $|0\rangle, |1\rangle, |2\rangle $ corresponding
to antiparallel, perpendicular and parallel orientation
of particle spin with respect ot the $z$ axis.
This means, in particular, that we put s=1 in formula (\ref{fid}).

For our purposes let us introduce mixed separable states:
\begin{eqnarray}
\sigma_{+}={1 \over 3}(|0\rangle|1\rangle \langle 0| \langle 1|
+ |1\rangle|2\rangle \langle 1| \langle 2|+
|2\rangle|0\rangle \langle 2| \langle 0|) \nonumber \\
\sigma_{-}={1 \over 3}(|1\rangle|0\rangle \langle 1| \langle 0|
+ |2\rangle|1\rangle \langle 2| \langle 1|+
|0\rangle|2\rangle \langle 0| \langle 2|)
\label{mix}
\end{eqnarray}

Suppose now that Alice nad Bob share {\it single}
pair of spin-1 particles in the free entangled
mixed state
\begin{equation}
\varrho_{free}=\varrho(F)\equiv F|\Psi_{+}\rangle \langle \Psi_{+}| +
(1-F)\sigma_{+}, \ \ 0<F<1
\label{source}
\end{equation}
In fact, it it is easy to see that the state is free
entangled. Namely after action of the
local projections $(|0\rangle \langle 0| +|1\rangle\langle 1|)\otimes
(|0\rangle\langle 0| +|1\rangle \langle 1|)$
we get inseparable $2 \times 2$ state. Such a result is known
\cite{bound} to be sufficient for distillability of entanglement from
the state, hence for the fact that
$\varrho(F)$ contains free entanglement.
On the other hand it can be shown \cite{dowod} that the state can never be quasi-distilled
noncollectively i.e. no QLCC performed on one pair in $\varrho(F)$ can increase its
fidelity upon some $C$
(we do not present the proof here, as it is a rather technical task
and requires some new approach then the ones applied so far).

Suppose, however, that apart from the state $\varrho(F)$ Alice and Bob
have a large number of pairs of  particles in the following state \cite{St}:
\begin{equation}
\sigma_{\alpha}=\frac{2}{7}|\Psi_{+}\rangle \langle \Psi_{+}|
+\frac{\alpha}{7} \sigma_{+}+
\frac{5-\alpha}{7} \sigma_{-}
\label{target}
\end{equation}
Those states admit simple characterization  with
respect to the parameter $2 \leq \alpha \leq 5$ :
\begin{eqnarray}
\sigma_{\alpha} \ \ \mbox{is} \
\left\{ \begin{array}{c}
\mbox{separable} \ \ \mbox{for} \ \ 2 \leq \alpha \leq 3 \\
\mbox{bound}  \ \ \mbox{entangled } \ \ \mbox{for} \ \ 3 < \alpha \leq 4 \\
\mbox{free} \ \ \mbox{entangled }  \ \ \mbox{for} \ \  4 < \alpha \leq 5  \\
\end{array} \right.
\end{eqnarray}
Let us briefly justify the above characterisation.
It is easy to point out separability of $\sigma_{\alpha}$
states for first region of parameter $2\leq \alpha \leq 3$. In fact then
it can be written as a mixture of
separable states (recall that separable states form the convex set)
$\sigma_{\alpha}=\frac{6}{7}\varrho_{1}
+\frac{\alpha-2}{7} \sigma_{+}+ \frac{3-\alpha}{7} \sigma_{-}$.
Here $\varrho_1=(|\Psi_{+}\rangle \langle \Psi_{+}|+
\sigma_{+}+\sigma_{-})/3$, which has been explicitly represented as
mixture of product states in Ref. \cite{tran}.
It is even more easy to find the free entanglement
of $\sigma_{\alpha}$ in the last region $4<\alpha\leq 5$,
as it can be done in the same way as for the
states (\ref{source}).
For intermediate region $3 < \alpha \leq 4 $ direct
calculation shows that $\sigma_{\alpha}$ satisfies
Peres separability criterion of positive partial
transposition \cite{Peres}. Nevertheless in this case the state
is inseparable,
and then, as shown in Ref. \cite{bound}, it is bound entangled.
Here, instead of direct
proving  of this inseparability, we will rather
show that such a states can produce
the effect which cannot come from any
separable state.
Namely we shall show that if only the states $\sigma_{\alpha}$
Alice and Bob share have
$3<\alpha\leq 4$ then they can quasi-distill the state $\varrho_{free}$.
Note, that it would {\it not} be possible
if the state were separable. Indeed, any usage of separable state
together with QLCC action 
could be interpreted as some new QLCC action {\it alone}, since
the separable state itself can be produced by means of some QLCC operation.
However, as it was mentioned before, {\it no} QLCC
on single pair in state (\ref{source}) can quasi-distill it.
Thus, the possibility of quasi-distillation of a
single pair $\varrho(F)$ with help
of state $\sigma_{\alpha}$, $3< \alpha \leq 4$
will be at the same time the proof that
the latter is bound entangled.
Note that any initial supply of BE states, if represents
the only entanglement in the process,
cannot be quasi-distilled \cite{be_fidelity_bound}.

Consider now the protocol of quasi-distillation.
Recall that Alice and Bob share one pair in the
FE state  (\ref{source}) and large supply of pairs
in BE states (\ref{target}).
They can procees repeating the folowing two
step procedure which is, in practice, the
direct $3 \times 3$ analogue of the one
used in destillation of entanglement \cite{d1,my1,xor}:

\vskip0.3cm
(i) They take the free entangled pair in the state $\varrho_{free}(F)$ and
one of the pairs being in the state $\sigma_{\alpha}$.
They perform the bilateral XOR operation
$U_{BXOR}=U_{XOR}\otimes U_{XOR}$, each of them treating the member
of free (bound) entangled pair as a source (target). Recall here that
the unitary XOR gate introduced in \cite{d1}, and used in generalised form in
\cite{xor,gott} is defined as
\begin{equation}
U_{XOR} | a \rangle | b \rangle= | a \rangle | b \oplus a \rangle, \ \
b \oplus a = (b + a){\rm mod} N
\end{equation}
where initial state $| a \rangle$  ($| b \rangle$ )
corresponds to source (target) state.
\vskip0.3cm

(ii) After that Alice and Bob measure the in their laboratories
the $z$-axis  spin components of the members of source pair.
Then they compare their results via phone. If the results
are the same they discard only target pair,
coming back with the, as we shall see, improved
source pair to the first step (i).
If the compared results appear to be different
they have to discard both pairs and
then the trial of improvement of $F$ fails.
\vskip0.3cm

By virtue of high symetry of the states (\ref{target}), (\ref{source})
it is easy to see that, conditioned that Alice
and Bob get the same results of their measurement
in step (ii), the above protocol leads with nonzero probability
\begin{equation}
P_{F \rightarrow F'} =\frac{2F + (1-F)(5 - \alpha) }{7}, \ \
\end{equation}
to the tranformation $\varrho(F) \rightarrow \varrho(F')$
where the improved fidelty $F'$ amounts to
\begin{equation}
F'(F)=
\frac{2F}{2F + (1-F)(5-\alpha)}
\label{func}
\end{equation}

If only $\alpha $ is greater than
$3$ the above continuous function
of F exeeds the value of F on the whole region $(0,1)$.
Thus the succesfull repeating of the steps (i-ii) produces
the sequence of source fidelities $F_{n} \rightarrow 1 $.
The probability of achieving any fidelity $F_{n}$
is $P_{n}=(P_{F \rightarrow F'})^{n}$ hence it is {\it nonzero}
for any $n$.
Thus all the states (\ref{target}) with $3<\alpha \leq 5$ allow us
to quasi-distill
state (\ref{source}). In particular the effect holds
for region $3<\alpha \leq 4$ confirming
that the target state (\ref{target}) is inseparable,
hence bound entangled in this region.
On the contrary for the region $2\leq \alpha \leq 3$ the iteration
of the formula (\ref{func})
decreases fidelity. This dramatic qualitatively
change reflects the fact
that then Alice's and Bob's large supply of pairs
is in separable states which,
as it was indicated before, cannot
help to quasi-distill pair in state (\ref{source}).
It is remarkable result as it shows that seemingly
usless bound entanglement can be, in a sense,
pumped into single pair of free entangled particles.
We expect similar effect for other bound entangled states like those
introduced in Ref. \cite{tran}.

Let us discuss the physical meaning
of the result. First we shall point out  an interesting connection
of the result with  the special kind of quantum communication
which is  teleportation. Recall that any quantum state of
composite system $\varrho$ can be regarded as a channel
in the process of the teleportation. The idea is that
Alice posses one particle in unknown state $\psi$
and one member of pair being in the state $\varrho$.
Bob posses another member of the pair.
After performing some deliberately chosen QLCC operation
Bob finds his particle in the state
resambling, at least to some degree, the initial
unknow state $\psi$ of Alice particle.
The fidelity of transmission of the state
is measured by the transfer fidelity
$f=\overline{\langle \psi| \Lambda_{\varrho}(\psi)| \psi \rangle} $
where $\Lambda_{\varrho}(\psi)$ represents Bob's particle
state after the whole procedure \cite{Pop} and the bar stands for average
over all possible input states $\psi$.
If the state $\varrho$ which forms  the quantum teleportation channel is
the maximally entangled state, then optimally
chosen QLCC guarantees $\Lambda_{\varrho}(\psi)=|\psi\rangle \langle \psi|$
and the transfer fidelity $f$ is equal to unity.
However in general, $f$ can be lesser than 1.
One can prove two simple connections between quantity $F$
and fidelity $f$ of teleportation
transfer.
Namely \cite{dowod}: ($\rm i'$) if
$F$ of a state $\varrho_{out}$ obtained form $\varrho$
by any QLCC operation is bounded by $C(\varrho)<1$
than $f$ of any teleporting procedure
through the new state $\varrho_{out}$
is certainly bounded by some other constant $d(\varrho)<1$;
($\rm ii'$) if for some
family of states $\varrho(F_{n})$ the fidelity
$F_{n}$ converges to 1
then under original teleportation procedure
\cite{Bennett_tel} the teleportation fidelity $f_{n}$
through $\varrho(F_{n})$ also converges to 1.

Let us now consider our result
in the context of these two facts.
For given $\varrho(F)$ we know that no QLCC can increase $F$
over some threshold $C(\varrho(F))<1$.
Hence according to (${\rm ii}'$) the teleportation transfer
fidelity $f$ after any QLCC operation is also bounded
by some $d(\varrho)<1$.
Suppose now that Alice wants to teleport
unknown spin-1 state $|\phi \rangle$
to Bob but only if she {\it is sure} that the state
is teleported with fidelity $f$ better than
some fixed bound $f_{r}$ satisfying $d(\varrho)<f_{r}<1$.
If the FE state (\ref{source})
is {\it the only} entangled state
shared by her and Bob then she will {\it never} decide to teleport,
as her requirement cannot be satisfied.
However,  according to the results of iteration of
scheme (i-ii) and the item (${\rm i}'$),
if apart from $\varrho_{free}$ Alice and Bob share a lot of
BE states (\ref{target})
then still  there is some nonzero chance that,
after some LQCC operations Alice can teleport,
being {\it sure} that her transmission has the required fidelity
$f>f_{r}$. Thus we see that bound entanglement
can lead to qualitative improvement of the
processes of quantum communication.

From the formula (\ref{func}) it follows that
the bound entanglement contained in the target state
($3< \alpha \leq 4$) behaves  qualitatively {\it in the same way}
as a free entanglement $(3< \alpha \leq 4)$. Moreover the effect would
not hold if only the single source pair were in bound entangled state.
But even if it is free entangled but alone, quasi-distillation
will not succeed. In contrast,  the  interaction between with large number of
BE states allows to make its fidelity arbitrary
close to 1.

A way of interpreting the results presented above is suggested by
entanglement-energy analogy \cite{bound}. Namely the situation is
somewhat similar to the processes 
which need an initial supply of some amount of energy to be run.
Here the role of the extra initial energy is played by the single
free entangled pair, which is allows to run the process of drawing
entanglement from the BE pairs.

In Ref. \cite{bound} the analogy entanglement-energy  was
stated quantitatively, where the analogue of useful
(free) energy was the distillable entanglement $D$.
Recall that the distillable entanglement
$D(\varrho)$ denotes the
maximal number of singlet pairs per input pair which can
be produced by means of QLCC operations
from large number of pairs in the state $\varrho$.
Now we can expect some other effect
being in more strict analogy with energy exchange
processes. Namely, we expect
that distillable entanglement \cite{huge} $D(\varrho)$
may be {\it non-additive}. In fact, by definition \cite{bound},
for any FE state $\varrho_{free}$ one has $D(\varrho_{free}) > 0$
while for BE ones $D(\varrho_{bound})=0$.
Note that the presented quasi-distillation scheme involves
some kind of entanglement transfer from BE pairs into the FE one.
It suggests that we may have
$D(\varrho_{free} \otimes \varrho_{bound})>
D(\varrho_{free})$. But the latter is simply
the sum $D(\varrho_{free}) + D(\varrho_{bound})$
as the last term vanish by definition.
This would really mimic a strange algebra
in which $0 + 1$ would be  greater than $1$.
Then the bound entanglement which
is {\it not distillable at all if alone} could be
distillable {\it through} free entanglement:
the latter would be the window allowing to liberate
the former. In terms of the mentioned analogy, the bound entanglement would
perform for us useful informational work, if supported by, perhaps small supply
free entanglement. Then, the role of the latter would be to activate the
bound one.

Even more probable effect strongly suggested by
the present results is the following.
Suppose that we enrich the actions Alice and Bob are conventionally
allowed to do. Namely, apart from performing local quantum operations and
classical communication, we allow them to share publicly any amount of
bound entangled pairs. Now, what have shown in this paper is that
the new class of operations (call it LQCC+BE) is significantly more powerful
than the LQCC operations alone.
Now one expects that the distillable entanglement
within this new paradigm can be strictly greater than the conventional one
i.e. we would have $D_{LQCC+BE}>D_{LQCC}$.

Finally, note that our discussion benefit from two opposite points
of view. In one of them we treat the bound entanglement as some supplement
which helps to handle with the free one, while in the other one,
the basis is bound entanglment, while the free one is only to activate it.
We believe that both perspectives will be useful for further
inverstigation of the role of bound entanglement in quantum information
theory.

M. H. and P. H. gratefully acknowledge the support
from Foundation for Polish Science.


\begin{references}
\bibitem{be_fidelity_bound}
Indeed, from however large supply of BE pairs one cannot produce
a spin-s pair of fidelity exceeding $1\over 2s+1$ because states
with such fidelity are FE (as shown in Ref. \cite{xor}).
\end{references}
\end{document}